\documentclass[11pt]{article}
\usepackage{amsmath}
\usepackage{graphicx,epstopdf}
\usepackage{cite}
\usepackage[font={small}]{caption}
\usepackage{xcolor}
\usepackage{graphicx}
\usepackage{tikz}
\def\checkmark{\tikz\fill[scale=0.4](0,.35) -- (.25,0) -- (1,.7) -- (.25,.15) -- cycle;}

\catcode`@=12
\usepackage[colorlinks=true,
            linkcolor=red,
            urlcolor=green,
            citecolor=blue]{hyperref} 
\def\beq{\begin{equation}}
\def\eeq{\end{equation}}
\def\bea{\begin{eqnarray}}
\def\eea{\end{eqnarray}}
\makeatletter
\@addtoreset{equation}{section}

\begin{document}
\title{\LARGE \bf Probing texture zeros with scaling ansatz in inverse seesaw }
\author{{\bf  Ambar Ghosal\footnote{ambar.ghosal@saha.ac.in}, Rome Samanta\footnote{rome.samanta@saha.ac.in}, }\\
  Saha Institute of Nuclear Physics, 1/AF Bidhannagar,
  Kolkata 700064, India \\
 }
\maketitle
\begin{abstract}
We investigate neutrino mass matrix phenomenology involving scaling ansatz and 
texture zeros adhering inverse seesaw mechanism. It is seen that four is the 
maximum number of zeros in $m_D$ and $\mu$ to obtain viable phenomenology. 
Depending upon the generic nature of the effective neutrino mass matrices we 
classify all the emerged matrices in four categories. One of them is ruled out  phenomenologically due to inappropriate value of reactor mixing angle after 
breaking of the scaling ansatz. The mass ordering is inverted in all cases. One of the distinguishable feature of all these categories is the vanishingly small value of CP violation measure $J_{CP}$ due to small value of $\delta_{CP}$. Thus those categories will be ruled out if CP violation is observed in the leptonic sector in future experiments.
\end{abstract}
\newpage
 \section{Introduction}
 Among the variants of the seesaw mechanism, inverse seesaw \cite{Mohapatra:1986bd,Bernabeu:1987gr,Mohapatra:1986aw,Schechter:1981cv,Palcu:2014aza,Schechter:1980gr
 ,Fraser:2014yha,Hettmansperger:2011bt,Adhikary:2013mfa,Law:2013gma,Dev:2012sg,Malinsky:2009dh,Hirsch:2009mx} stands out as an attractive one due to its characteristic feature of generation of small neutrino mass without invoking high energy scale in the theory. Although to realize such feature one has 
to pay the price in terms of incorporation of additional singlet fermions, nevertheless, in different GUT models accommodation of such type of neutral fermions are natural. Furthermore, such mechanism appeals to the foreseeable collider experiments to be testified due to its unique signature. The $9\times9$ neutrino mass matrix in this mechanism is written as 
 \bea
 m_\nu &=& \begin{pmatrix}
 0&m_D&0\\m_D^T&0&M_{RS}\\
 0&M_{RS}^T&\mu\\
 \end{pmatrix}
 \eea
 with the choice of basis $(\nu_L,\nu_R^c,S_L)$. 
The three matrices appear in $m_\nu$ are $m_D$, $M_{RS}$ and $\mu$ among them $m_D$ and $M_{RS}$ are Dirac type whereas $\mu$ is Majorana type mass matrix. After diagonalization, the low energy effective neutrino mass comes out as  
\bea
m_\nu &=& m_D M_{RS}^{-1}\mu ( m_D M_{RS}^{-1})^T \nonumber \\
 &=&F\mu F^T\label{1} 
\eea
where $F=m_D M_{RS}^{-1}$. Such definition resembles the above formula as a conventional type-I seesaw expression of $m_\nu$. However, $m_{\nu}$ contains large number of parameters and it is possible to fit them with neutrino oscillation experimental data \cite{Forero:2014bxa,GonzalezGarcia:2012sz,Tortola:2012te} (but the predictability is less). Our goal in this work is to find out a phenomenologically viable texture of $m_D$ and $\mu$ with minimum number of parameters or equivalently maximum number of zeros. We bring together two theoretical ideas to find out a minimal texture and they are \\
i) Scaling ansatz\cite{
  sc1,sc2,Blum:2007qm,Obara:2007nb,Damanik:2007yg,Goswami:2008rt,Grimus:2004cj,Berger:2006zw,sc3,Dev:2013esa,
  Adhikary:2012kb}, \\
ii) Texture Zeros\cite{Frampton:2002yf,Whisnant:2014yca,Ludl:2014axa,Grimus:2013qea,Liao:2013saa,Fritzsch:2011qv,Merle:2006du,Wang:2013woa,Wang:2014dka,Lavoura:2004tu,Kageyama:2002zw,Wang:2014vua,4zero1,Choubey:2008tb,Chakraborty:2014hfa,4zero2,4zero3,4zero4}.\\
\vskip 0.1in
\noindent
At the outset of the analysis, we choose a basis where the charged lepton mass matrix ($m_E$) and  $M_{RS}$ are diagonal along with texture zeros in $m_D$ and $\mu$ matrices. We also start by assuming the scaling property in the elements of $m_D$ and $\mu$ to reduce the number of relevant matrices. Although, we are not addressing the explicit origin of such choice of matrices, however, qualitatively we can assume that this can be achieved due to some flavour symmetry \cite{Grimus:2004hf} which is required to make certain that the texture zeros appear in $m_D$ and $\mu$ are in the same basis in which $m_E$ and  $M_{RS}$ are diagonal. We restrict ourselves within the frame work of $SU(2)_L \times U(1)_Y$ gauge group however, explicit realization of such scheme obviously more elusive which will be studied elsewhere. 
\section{Scaling property and texture zeros}
We consider scaling property between the second and third row of $m_D$ matrix and the same for $\mu$ matrix also. Explicitly the relationships are written as 
\bea
\frac{(m_D)_{2i}}{(m_D)_{3i}} &=& k_1\label{2.1}\\
\frac{(\mu)_{2i}}{(\mu)_{3i}} &=& k_2\label{2.2}
\eea
where $i=1,2,3$ is the column index. We would like to mention that although we have considered different scale factors for $m_D$ and $\mu$ matrices, however, the effective $m_\nu$ is still scale invariant and leads to $\theta_{13}=0$. Thus, it is obvious to further break the scaling ansatz. In order to generate nonzero $\theta_{13}$ it is necessary to break the ansatz in $m_D$ since, breaking in $\mu$ does not affect the generation of nonzero $\theta_{13}$ although in some cases it provides $m_3\neq 0$. In our scheme texture zero format is robust and it remains intact while the scaling ansatz is explicitly broken. Such a scenario can be realized by considering the scaling ansatz and texture zeros to have a different origin.\\
\noindent
Another point is to be noted that, since the $\mu$ matrix is complex symmetric whereas $m_D$ is asymmetric, the scale factor considered in $\mu$ matrix is different from that of $m_D$ to keep the row wise invariance as dictated by Eqn.(\ref{2.1}) (for $m_D$), and Eqn.(\ref{2.2}) (for $\mu$). Finally, since the texture of $M_{RS}$ matrix is diagonal it is not possible to accommodate scaling ansatz considered in the present scheme.\\
\noindent
Let us now turn to further constrain the matrices assuming zeros in different entries. Since, in our present scheme the matrix $M_{RS}$ is diagonal, we constrain the other two matrices. We start with the maximal zero textures with scaling ansatz of general $3\times 3$ matrices and list different cases  systematically in \textbf{Table \ref{t1}}.\\
\begin{table}[!h]
\caption{Texture zeros with scaling ansatz of a general $3\times3$ matrix} \label{t1}
\begin{center}
\begin{tabular}{|c|c|c|}
\hline
\multicolumn{3}{|c|}{{\bf $7$ zero texture}}\\
\hline
$m_1^7=\begin{pmatrix}
0&0&0\\
k_1 c_1&0&0\\
c_1&0&0\\
\end{pmatrix}$ & 
$m_2^7=\begin{pmatrix}
0&0&0\\
0&k_1 c_2&0\\
0&c_2&0\\
\end{pmatrix}$ &
$m_3^7=\begin{pmatrix}
0&0&0\\
0&0&k_1 c_3\\
0&0&c_3\\
\end{pmatrix}$\\
 \hline
\end{tabular}
\end{center}
\end{table}
\begin{center}
\begin{tabular}{|c|c|c|}
\hline
\multicolumn{3}{|c|}{{\bf $6$ zero texture}}\\
\hline
$m_1^6=\begin{pmatrix}
d_1&0&0\\
k_1 c_1&0&0\\
c_1&0&0\\
\end{pmatrix}$ & 
$m_2^6=\begin{pmatrix}
0&d_2&0\\
k_1 c_1&0&0\\
c_1&0&0\\
\end{pmatrix}$ &
$m_3^6=\begin{pmatrix}
0&0&d_3\\
k_1 c_1&0&0\\
c_1&0&0\\
\end{pmatrix}$\\
 
\hline
$m_4^6=\begin{pmatrix}
d_1&0&0\\
0&k_1 c_2&0\\
0&c_2&0\\
\end{pmatrix}$ &
$m_5^6=\begin{pmatrix}
0&d_2&0\\
0&k_1 c_2&0\\
0&c_2&0\\
\end{pmatrix}$ &
$m_6^6=\begin{pmatrix}
0&0&d_3\\
0&k_1 c_2&0\\
0&c_2&0\\
\end{pmatrix}$\\
\hline
$m_7^6=\begin{pmatrix}
d_1&0&0\\
0&0&k_1 c_3\\
0&0&c_3\\
\end{pmatrix}$ &
$m_8^6=\begin{pmatrix}
0&d_2&0\\
0&0&k_1 c_3\\
0&0&c_3\\
\end{pmatrix}$ &
$m_9^6=\begin{pmatrix}
0&0&d_3\\
0&0&k_1 c_3\\
0&0&c_3\\
\end{pmatrix}$ \\
\hline
\multicolumn{3}{|c|}{{\bf $5$ zero texture}}\\
\hline
$m_1^5=\begin{pmatrix}
0&0&0\\
k_1 c_1&k_1c_2&0\\
c_1&c_2&0\\
\end{pmatrix}$ &
$m_2^5=\begin{pmatrix}
0&0&0\\
k_1 c_1&0&k_1 c_3\\
c_1&0&c_3\\
\end{pmatrix}$&
$m_3^5=\begin{pmatrix}
0&0&0\\
0&k_1 c_1&k_1c_3\\
0&c_1&c_3\\
\end{pmatrix}$\\
\hline
$m_4^5=\begin{pmatrix}
d_1&d_2&0\\
k_1 c_1&0&0\\
c_1&0&0\\
\end{pmatrix}$ & 
$m_5^5=\begin{pmatrix}
0&d_2&d_3\\
k_1 c_1&0&0\\
c_1&0&0\\
\end{pmatrix}$ &
$m_6^5=\begin{pmatrix}
d_1&0&d_3\\
k_1 c_1&0&0\\
c_1&0&0\\
\end{pmatrix}$\\
 
\hline
$m_7^5=\begin{pmatrix}
d_1&d_2&0\\
0&k_1 c_2&0\\
0&c_2&0\\
\end{pmatrix}$ &
$m_8^5=\begin{pmatrix}
0&d_2&d_3\\
0&k_1 c_2&0\\
0&c_2&0\\
\end{pmatrix}$ &
$m_9^5=\begin{pmatrix}
d_1&0&d_3\\
0&k_1 c_2&0\\
0&c_2&0\\
\end{pmatrix}$\\
\hline
$m_{10}^5=\begin{pmatrix}
d_1&d_2&0\\
0&0&k_1 c_3\\
0&0&c_3\\
\end{pmatrix}$ &
$m_{11}^5=\begin{pmatrix}
0&d_2&d_3\\
0&0&k_1 c_3\\
0&0&c_3\\
\end{pmatrix}$ &
$m_{12}^5=\begin{pmatrix}
d_1&0&d_3\\
0&0&k_1 c_3\\
0&0&c_3\\
\end{pmatrix}$ \\
\hline
\multicolumn{3}{|c|}{{\bf $4$ zero texture}}\\
\hline
$m_1^4=\begin{pmatrix}
d_1&0&0\\
0&k _1c_2&k_1 c_3\\
0&c_2&c_3\\
\end{pmatrix}$ & 
$m_2^4=\begin{pmatrix}
0&d_2&0\\
0&k_1 c_2&k_1 c_3\\
0&c_2&c_3\\
\end{pmatrix}$ & 
$m_3^4=\begin{pmatrix}
0&0&d_3\\
0&k_1 c_2&k_1 c_3\\
0&c_2&c_3\\
\end{pmatrix}$\\
\hline
$m_4^4=\begin{pmatrix}
d_1&0&0\\
k_1 c_1&0&k_1 c_3\\
c_1&0&c_3\\
\end{pmatrix}$ & 
$m_5^4=\begin{pmatrix}
0&d_2&0\\
k_1c_1&0&k_1 c_3\\
c_1&0&c_3\\
\end{pmatrix}$ & 
$m_6^4=\begin{pmatrix}
0&0&d_3\\
k_1 c_1&0&k_1 c_3\\
c_1&0&c_3\\
\end{pmatrix}$\\
\hline
$m_7^4=\begin{pmatrix}
d_1&0&0\\
k_1 c_1&k_1 c_2&0\\
c_1&c_2&0\\
\end{pmatrix}$ & 
$m_8^4=\begin{pmatrix}
0&d_2&0\\
k_1 c_1&k_1 c_2&0\\
c_1&c_2&0\\
\end{pmatrix}$ & 
$m_9^4•=\begin{pmatrix}
0&0&d_3\\
k_1 c_1&k_1 c_2&0\\
c_1&c_2&0\\
\end{pmatrix}$\\
\hline
$m_{10}^4=\begin{pmatrix}
d_1&d_2&d_3\\
k_1 c_1&0&0\\
c_1&0&0\\
\end{pmatrix}$ & 
$m_{11}^4=\begin{pmatrix}
d_1&d_2&d_3\\
0&k_1 c_2&0\\
0&c_2&0\\
\end{pmatrix}$ & 
$m_{12}^4=\begin{pmatrix}
d_1&d_2&d_3\\
0&0&k_1 c_3\\
0&0&c_3\\
\end{pmatrix}$\\
\hline
\end{tabular}
\end{center}
\noindent
We consider all the matrices\footnote{From now on we use $m^n$ 
as a mass matrix where $n(=4,5,6,7)$ is the number of zeros in that matrix.} 
listed in \textbf{Table \ref{t1}} as the Dirac type matrices($m_D$). As 
the lepton number violating mass matrix $\mu$ is complex symmetric, therefore, 
the maximal number of zeros with scaling invariance is 5. Therefore, only $m_3^5$ and $m_5^5$ type matrices can be made complex symmetric with the scaling property and are shown   in \textbf{Table \ref{t2}} where they are renamed as $\mu_1^5$ and $\mu_2^5$ with a different scale factor $k_2$.\\

\begin{table}[!h]
\caption{ Maximal zero texture of $\mu$ matrix}\label{t2}
\begin{center}
\begin{tabular}{|c|c|}
\hline
$\mu_1^5=\begin{pmatrix}
0&0&0\\
0&k_2^2s_3&k_2 s_3\\
0&k_2 s_3&s_3\\
\end{pmatrix}$ & 
$\mu_2^5=\begin{pmatrix}
0&k_2s_3&s_3\\
k_2s_3&0&0\\
s_3&0&0\\
\end{pmatrix}$ \\ 
\hline
\end{tabular}
\end{center}
\end{table}
Now using Eqn.(\ref{1}) we can construct $m_\nu$ and it is found that all the mass matrices constructed out of these matrices are not suitable to satisfy the neutrino oscillation data. The reason goes as follows:\\

\noindent
\textbf{Case A:} $m_D$ (7, 6 zero) + $\mu_1^5$, $\mu_2^5$ (5 zero):\\

\noindent
We can not generate nonzero $\theta_{13}$ by breaking the scaling ansatz 
because in this case all the structures of $m_D$ are scaling ansatz invariant. 
This can be understood in the following way: if we incorporate scaling ansatz 
breaking by $k_1^\prime\rightarrow  k_1(1+\epsilon)$ all the structures of 
$m_D$ are still invariant and $m_\nu$ matrix will still give $\theta_{13}=0$ as breaking of scaling in $\mu_1^5$ and $\mu_2^5$ play no role for the generation of nonzero  value of $\theta_{13}$. To generate nonzero $\theta_{13}$ it is necessary to break scaling ansatz in the Dirac sector.\\

\noindent
\textbf{Case B:} $m_D$ (5 zero) + $\mu_1^5$, $\mu_2^5$ (5 zero):\\

\noindent
The matrices in the last three rows ($m_4^5$ to $m_{12}^5$) of the 
`\textbf{5 zero texture}' part of \textbf{Table \ref{t1}} are ruled out due to 
the same reason as mentioned in \textbf{Case A} while, the matrices in the first row i.e. $m_1^5$, $m_2^5$ and $m_3^5$ give rise to the structure of $m_\nu$ as \\
\beq
A_1=\begin{pmatrix}
0&0&0\\
0&\ast&\ast\\
0&\ast&\ast\\
\end{pmatrix}\label{A1}
\eeq
where `$\ast$' represents some nonzero entries in $m_\nu$. This structure leads to complete disappearance of one generation. Moreover it has been shown in Ref. \cite{Frampton:2002yf} that if the number of independent zeros in an effective neutrino mass matrix ($m_\nu$) is $\geq 3$ it doesn't favour the oscillation data and hence, `$A_1$' type mass matrix is ruled out.\\

\noindent
\textbf{Case C:} $m_D$ (4 zero) + $\mu_1^5$ (5 zero):\\

\noindent
There are 12 $m_D$ matrices with 4 zero texture and they are designated as $m_1^4$,...$m_{12}^4$ in \textbf{Table \ref{t1}}. Due to the same reason as discussed in \textbf{Case A}, $m_{10}^4$, $m_{11}^4$ and $m_{12}^4$ are not considered. Furthermore, $m_\nu$ arises through $m_{1}^4$, $m_{4}^4$ and $m_{7}^4$ also correspond to the `$A_1$' type matrix (shown in Eqn.(\ref{A1})) and hence are also discarded. Finally, remaining six $m_D$ matrices $m_{2}^4$,  $m_{3}^4$, $m_{5}^4$, $m_{6}^4$, $m_{8}^4$ and $m_{9}^4$ lead to the structure of $m_\nu$ with two zero eigenvalues and obviously they are also neglected.\\

\noindent
\textbf{Case D:} $m_D$ (4 zero) + $\mu_2^5$ (5 zero):\\

\noindent
 In this case, for $m_2^4$ and $m_3^4$ the low energy mass matrix $m_\nu$ comes out as a null matrix while for $m_1^4$ the structure of $m_\nu$ is given by \beq
A_2=\begin{pmatrix}
0&\ast & \ast \\
\ast & 0&0\\
\ast &0 &0\\
\end{pmatrix}
\eeq
which is also neglected since the number of independent zeros $\geq 3$.\\
On the other hand rest of the $m_D$ matrices ( $m_4^4$ to $m_9^4$ ) correspond to the structure of $m_\nu$ as
\beq
A_3=\begin{pmatrix}
0&\ast & \ast \\
\ast & \ast &\ast \\
\ast & \ast &\ast\\
\end{pmatrix}.\label{rldout}
\eeq
Interestingly, a priori we cannot rule out the matrices of type $A_3$, however it is observed that  $m_\nu$ of this type fails to generate $\theta_{13}$ within the present experimental bound (details are mentioned in section (\ref{s1})). It is also observed that in this scheme to generate viable neutrino oscillation data, four zero texture of both $m_D$ and $\mu$ matrices are necessary. Therefore, now on we discuss extensively the four zero texture in both the sectors ( Dirac as well as Majorana sector ).
\section{4 zero texture}
There are 126 ways to choose 4 zeros out of 9 elements of a general $3\times3$ matrix. Hence there are $ 126 $ textures. Incorporation of scaling ansatz leads to a drastic reduction to only 12 textures as given in the \textbf{Table \ref{t1}}. In our chosen basis since $M_{RS}$ is taken as diagonal, therefore, the structure of $m_D$ leads to the same structure of $F$. On the other hand the lepton number violating mass matrix $\mu$ is complex symmetric and therefore from the matrices listed in \textbf{Table \ref{t1}}, only $m_1^4$ and $m_{10}^4$ type matrices are acceptable. We renamed those matrices as $\mu_1^4$ and $\mu_2^4$ and explicit structures of them are presented in \textbf{Table \ref{t3}}.\\

\begin{table}[!h]
\caption{Four zero texture of $\mu$ matrix}\label{t3}
\begin{center}
\begin{tabular}{|c|c|}
\hline
$\mu_1^4=\begin{pmatrix}
r_1&0&0\\
0&k_2^2s_3&k_2 s_3\\
0&k_2 s_3&s_3\\
\end{pmatrix}$ & 
$\mu_2^4=\begin{pmatrix}
r_1&k_2s_3&s_3\\
k_2s_3&0&0\\
s_3&0&0\\
\end{pmatrix}$ \\ 
\hline
\end{tabular}
\end{center}
\end{table}
\noindent
There are now $2\times12=24$ types of $m_\nu$ due to both the choices of 
$\mu$ matrices. We discriminate different types of $m_D$ matrices in the following way:\\
i) First of all, the texture $m_{10}^4$, $m_{11}^4$ and $m_{12}^4$ are always scaling ansatz invariant due to the same reason mentioned earlier in \textbf{Case A} and hence are all discarded.\\
Next the matrices $m_1^4$, $m_2^4$ and $m_3^4$ are also ruled out due to the following:\\
a) When $\mu_1^4$ matrix is taken to generate $m_\nu$ along with 
$m_1^4$, $m_2^4$ and $m_3^4$ as the Dirac matrices, then the structure of the 
effective $m_\nu$ appears such that, one generation is completely decoupled 
thus leading to two mixing angles zero for the matrix $m_1^4$ and two zero
eigenvalues when we consider $m_2^4$ and $m_3^4$ matrices.\\
b) In case of $\mu_2^4$ matrix, the form of $m_\nu$ for $m_1^4$ comes out as 
\bea
A_4&=&
\begin{pmatrix}
\ast & \ast & \ast\\
\ast &0&0\\
\ast &0 &0\\
\end{pmatrix}
\eea
which is phenomenologically ruled out and for other two matrices ($m_2^4$ and $m_3^4$) $m_\nu$ becomes a null matrix. For a compact view of the above analysis we present the ruled out and survived structures of $m_\nu$ symbolically in \textbf{Table \ref{sym}}.\\
\newpage
\begin{table}[!h]
\caption{ Compositions of the discarded and survived structures of $m_\nu$}\label{sym}
\begin{center}
\begin{tabular}{|p{1cm}|p{1cm}|p{1cm}|p{1cm}|p{1cm}|p{1cm}|p{1cm}|p{1cm}|p{1cm}|p{1cm}|p{1cm}|p{1cm}|p{1cm}|}
\cline{2-13}
\multicolumn{1}{c|}{}  & \multicolumn{12}{c|}{$m_D$} \\ \hline
$\mu$ & $m_1^4$ &$m_2^4$ & $m_3^4$ & $m_4^4$ & $m_5^4$ & $m_6^4$ & $m_7^4$ & $m_8^4$ & $m_9^4$ & $m_{10}^4$ & $m_{11}^4$& $m_{12}^4$\\
\hline
$ \mu_1^4 $ & $\times$ & $\times$ &$\times$ & $\checkmark$ & $\checkmark$ & $\checkmark$& $\checkmark$ & $\checkmark$ & $\checkmark$ & $\times$& $\times$ & $\times$\\
\hline
$\mu_2^4$&$\times$&$\times$&$\times$&$\checkmark$&$\checkmark$&$\checkmark$&$\checkmark$&$\checkmark$&$\checkmark$&$\times$&$\times$&$\times$\\
\hline
\end{tabular}
\end{center}
\end{table} 
\noindent
Thus we are left with same six textures of $m_D$ for both the choices of $\mu$ and they are renamed in \textbf{Table \ref{t4}} as $m_{D1}^4$, $m_{D2}^4$, ....
$m_{D_6}^4$.
\begin{table}[!h]
\caption{Four zero textures of the Dirac mass matrices} \label{t4}
 \begin{center}
 \begin{tabular}{|c|c|c|}
 \hline
$m_{D1}^4=\begin{pmatrix}
d_1&0&0\\
k_1 c_1&0&k_1 c_3\\
c_1&0&c_3\\
\end{pmatrix}$ & 
$m_{D2}^4=\begin{pmatrix}
0&d_2&0\\
k_1 c_1&0&k_1 c_3\\
c_1&0&c_3\\
\end{pmatrix}$ & 
$m_{D3}^4=\begin{pmatrix}
0&0&d_3\\
k_1 c_1&0&k_1 c_3\\
c_1&0&c_3\\
\end{pmatrix}$\\
\hline
$m_{D4}^4=\begin{pmatrix}
d_1&0&0\\
k_1c_1&k_1 c_2&0\\
c_1&c_2&0\\
\end{pmatrix}$ & 
$m_{D5}^4=\begin{pmatrix}
0&d_2&0\\
k_1 c_1&k_1 c_2&0\\
c_1&c_2&0\\
\end{pmatrix}$ & 
$m_{D6}^4=\begin{pmatrix}
0&0&d_3\\
k_1 c_1&k_1 c_2&0\\
c_1&c_2&0\\
\end{pmatrix}$\\
\hline
\end{tabular}
 \end{center}
 \end{table}
 
\noindent 
 Obviously, it is clear that the above analysis leads to altogether 12 effective $m_\nu$ matrices arising due to six $m_D$ ($m_{D1}^4$ to $m_{D6}^4$) and two $\mu$ ($\mu_1^4$ and $\mu_2^4$) matrices.
\section{ Parametrization}
Depending upon the composition of $m_D$ and $\mu$ we subdivided those 12 
$m_\nu$ matrices in four broad categories and each category is again 
separated in few cases and the decomposition is presented in 
\textbf{Table \ref{t5}} and \textbf{Table \ref {t6}}.\\
\begin{table}[!h]
\caption{Different Composition of $m_D$ and $\mu_1$ matrices to generate $m_\nu$.}\label{t5}
\begin{center}
\begin{tabular}{|p{1.5cm}|p{1.5cm}|p{1.5cm}|p{1.5cm}|p{1.5cm}|p{1.5cm}|p{1.5cm}|}
\cline{2-7}
\multicolumn{1}{c|}{}  & \multicolumn{2}{c|}{\textbf{Category A}} & \multicolumn{4}{c|}{\textbf{Category B}}\\ \hline
$\textbf{Matrices}$ &$I_A$&$II_A$&$I_B$&$II_B$&$III_B$&$IV_B$\\
\hline
$ m_D $ & $m_{D2}^4$ & $m_{D6}^4$ & $m_{D1}^4$ & $m_{D3}^4$ & $m_{D4}^4$ & $m_{D5}^4$\\
\hline
$\mu$ &$\mu_1^4$&$\mu_1^4$&$\mu_1^4$&$\mu_1^4$&$\mu_1^4$&$\mu_1^4$\\
\hline
\end{tabular}
\end{center}
\end{table}

\begin{table}[!h]
\caption{ Different Composition of $m_D$ and $\mu_2$ matrices to generate 
$m_\nu$.}\label{t6}
\begin{center}
\begin{tabular}{|p{1.5cm}|p{1.5cm}|p{1.5cm}|p{1.5cm}|p{1.5cm}|p{1.5cm}|p{1.5cm}|}
\cline{2-7}
\multicolumn{1}{c|}{}  & \multicolumn{2}{c|}{\textbf{Category C}} & \multicolumn{4}{c|}{\textbf{Category D}}\\ \hline
$\textbf{Matrices}$ &$I_C$&$II_C$&$I_D$&$II_D$&$III_D$&$IV_D$\\
\hline
$ m_D $ & $m_{D1}^4$ & $m_{D4}^4$ & $m_{D2}^4$ & $m_{D3}^4$ & $m_{D5}^4$ & $m_{D6}^4$\\
\hline
$\mu$ &$\mu_2^4$&$\mu_2^4$&$\mu_2^4$&$\mu_2^4$&$\mu_2^4$&$\mu_2^4$\\
\hline
\end{tabular}
\end{center}
\end{table}
\noindent
Throughout our analysis we consider the matrix $M_{RS}$ as 
\bea
M_{RS}&=&\begin{pmatrix}
p_1&0&0\\
0&p_2&0\\
0&0&p_3\\
\end{pmatrix}.
\eea
Following Eqn.(\ref{1}), the $m_\nu$ matrix arises in Category A and Category B can be written in a generic way as 
\bea
m_\nu^{AB} &= &m_0\begin{pmatrix}
1& k_1p & p\\
k_1p & k_1^2(q^2+p^2)& k_1(q^2+p^2)\\p& k_1(q^2+p^2)&(q^2+p^2)
\end{pmatrix}\label{4}
\eea
with the definition of parameters as following\\
\bea
\textbf{$Set$ $I_A:\;$}   m_0^\prime =\frac{d_3^2s_3}{p_3^2},p^\prime =  \frac{p_3 c_2}{p_2 d_3},q^\prime =\frac{c_1p_3}{d_3p_1}\sqrt{\frac{r_1}{s_3}},m_0 = m_0^\prime ,p=k_2p^\prime ,q=q^\prime  \nonumber \\
\textbf{$Set$ $II_A:\;$} m_0^\prime = \frac{d_2^2s_3}{p_2^2}, p^\prime =\frac{p_2 c_2}{p_3 d_2},q^\prime =\frac{c_1p_2}{d_2p_1}\sqrt{\frac{r_1}{s_1}},m_0=m_0^\prime k_2^2,p=\frac{p^\prime}{k_2},q=\frac{q^\prime}{k_2}\nonumber \\
\textbf{$Set$ $I_B:\;$} m_0^\prime = \frac{d_1^2r_1}{p_1^2},p^\prime =\frac{ c_1}{ d_1},q^\prime = \frac{c_3p_1}{d_1p_3}\sqrt{\frac{s_3}{r_1}},m_0=m_0^\prime , p=p^\prime ,q=q^\prime \nonumber \\
\textbf{$Set$ $II_B:\;$} m_0^\prime = \frac{d_3^2s_3}{p_3^2},p^\prime =\frac{ c_3}{ d_3},q^\prime =\frac{c_1p_3}{d_3p_1}\sqrt{\frac{r_1}{s_1}},m_0=m_0^\prime ,p=p^\prime ,q=q^\prime \nonumber \\
\textbf{$Set$ $III_B:\;$} m_0^\prime =\frac{d_1^2r_1}{p_1^2},p^\prime = \frac{ c_1}{ d_1},q^\prime =\frac{c_2p_1}{d_1p_2}\sqrt{\frac{s_3}{r_1}},m_0=m_0^\prime ,p = p^\prime ,q=k_2q^\prime \nonumber \\
\textbf{$Set$ $IV_B:\;$}  m_0^\prime =\frac{d_2^2s_3}{p_2^2}, p^\prime =\frac{ c_2}{d_2},q^\prime =\frac{c_1p_2}{d_2p_1}\sqrt{\frac{r_1}{s_1}},m_0=m_0^\prime k_2^2,p={p^\prime},q=\frac{q^\prime}{k_2}.\label{p1}
\eea
 Similarly the $m_\nu$ matrix arises in Category C can be written as 
\bea
m_\nu^{C}& =& m_0 \begin{pmatrix}
 1&k_1(p+q)&p+q\\
 k_1(p+q)&k_1^2(2pq+p^2)&k_1(2pq+p^2)\\
 p+q&k_1(2pq+p^2)&(2pq+p^2)\\
 \end{pmatrix}\label{7}
 \eea
with the following choice of parameters
\bea
\textbf{$Set$ $I_C:\;$} m_0^\prime =\frac{d_1^2r_1}{p_1^2•},p^\prime = \frac{c_1}{ d_1},q^\prime =\frac{c_2p_1•}{d_1p_2•}\sqrt{\frac{s_3}{r_1•}},m_0=m_0^\prime ,p=p^\prime ,q=k_2q^\prime \nonumber \\
\textbf{$Set$ $II_C:\;$} m_0^\prime =\frac{d_1^2r_1}{p_1^2•}, p^\prime =\frac{c_1}{ d_1},q^\prime =\frac{c_3p_1•}{d_1p_3•}\sqrt{\frac{s_3}{r_1•}},m_0=m_0^\prime ,p=p^\prime ,q=q^\prime . 
\eea
For Category D the effective $m_\nu$ comes out as 
\bea
m_\nu^{D} &=&m_0\begin{pmatrix}
0&k_1p&p\\
k_1p&k_1^2(q^2+2rp)&k_1(q^2+2rp)\\
p&k_1(q^2+2rp)&(q^2+2rp)
\end{pmatrix}\eea
with the definition of parameters as
\bea
\textbf{$Set$ $I_D:\;$} m_0^\prime =\frac{d_2^2r_1}{p_1^2•},p^\prime = \frac{ c_1p_1s_3}{ d_2p_2r_1}, q^\prime =\frac{c_1•}{d_2•},r^\prime =\frac{c_3}{d_2•},m_0=m_0^\prime ,p=k_2p^\prime ,q=q^\prime ,r=r^\prime \nonumber \\
\textbf{$Set$ $II_D:\;$} m_0^\prime =\frac{d_3^2r_1}{p_1^2•},p^\prime = \frac{ c_1p_1s_3}{ d_3p_3r_1},q^\prime =\frac{c_1}{d_3•},r^\prime =\frac{c_2}{•d_3},m_0=m_0^\prime ,p = p^\prime ,q=q^\prime r=k_2 r^\prime \nonumber \\
\textbf{$Set$ $III_D:\;$}  m_0^\prime =\frac{ c_1p_1s_3}{ d_3p_3r_1},p^\prime = \frac{ c_1}{ d_1}, q^\prime =\frac{c_1}{d_3•}, r^\prime =\frac{c_3}{•d_3},m_0=m_0^\prime ,p =p^\prime ,q=k_2q^\prime ,r=r^\prime \nonumber\\
\textbf{$Set$ $IV_D:\;$} m_0^\prime =\frac{d_2^2r_1}{p_1^2•},p^\prime = \frac{ c_1p_1s_3}{ d_2p_2r_1}, q^\prime =\frac{c_1•}{d_2•},r^\prime =\frac{c_2}{d_2•},m_0=m_0^\prime ,p=k_2{p^\prime} ,q=q^\prime ,r =r^\prime 
\eea
and in general, we consider all the parameters $m_0$, $k_1$, $p$, $r$ and $q$ are complex.
 \section{Phase Rotation}
As mentioned earlier, all the parameters of $m_\nu$ are complex and therefore we can rephase $m_\nu$ by a phase rotation to remove the redundant phases. Here, we systematically study the phase rotation for each category.\\
\noindent
\textbf{Category A,B}\\ 
\noindent
The Majorana type mass matrix $m_\nu$ can be rotated in phase space through 
\bea
m_\nu^{\prime AB}= P^T m_\nu^{AB} P\label{pp}
\eea
 where $P$ is a diagonal phase matrix and is given by $P=diag(e^{i\Phi_1},e^{i\Phi_2},e^{i\Phi_3})$.\\
 Redefining the  parameters of $m_\nu$ as \bea
 m_0\rightarrow m_0e^{i\alpha_m}, p\rightarrow pe^{i\theta_p}, q\rightarrow qe^{i\theta_q}, k_1\rightarrow k_1e^{i\theta_1}\label{prmtr}\eea
  with 
  \bea \Phi_1=-\frac{\alpha_m}{2•},\Phi_2=-(\theta_1+\theta_p+\frac{\alpha_m}{2•}), \Phi_3=-(\theta_p+\frac{\alpha_m}{2})\label{phi}\eea
 the phase rotated $m_\nu^{\prime AB}$ appears as \bea
  m_\nu^{\prime AB}&=& m_0\begin{pmatrix}
1& k_1p & p\\
k_1p & k_1^2(q^2e^{i\theta}+p^2)& k_1(q^2e^{i\theta}+p^2)\\p& k_1(q^2e^{i\theta}+p^2)&(q^2e^{i\theta}+p^2)
\end{pmatrix}\label{8}
\eea
where $\theta=2(\theta_q-\theta_p)$ and all the parameters $m_0,p,q$ and $k_1$ are real. Thus there is only a single phase parameter in $ m_\nu^{\prime AB}$.\\
\noindent
\textbf{Category C}\\ 
\noindent
In a similar way, the mass matrix of Category C can be rephased as 
\bea
m_\nu^{\prime C} &=& m_0 \begin{pmatrix}
 1&k_1(p+qe^{i\theta})&p+qe^{i\theta}\\
 k_1(p+qe^{i\theta})&k_1^2(2pqe^{i\theta}+p^2)&k_1(2pqe^{i\theta}+p^2)\\
 p+qe^{i\theta}&k_1(2pqe^{i\theta}+p^2)&(2pqe^{i\theta}+p^2)\\
 \end{pmatrix}\label{9}
 \eea
with the same set of redefined parameters as mentioned in Eqn.(\ref{prmtr})and (\ref{phi}) and the diagonal phase matrix mentioned in the previous case with $\theta=\theta_q-\theta_p$.\\
\noindent
\textbf{Category D} \\
For this category the rephased mass matrix comes out as
\bea
m_\nu^{\prime D} &=&m_0\begin{pmatrix}
0&k_1p&p\\
k_1p&k_1^2(q^2e^{i\alpha}+2rpe^{i\beta})&k_1(q^2e^{i\alpha}+2rpe^{i\beta})\\
p&k_1(q^2e^{i\alpha}+2rpe^{i\beta})&(q^2e^{i\alpha}+2rpe^{i\beta})
\end{pmatrix}\label{10}
\eea
with $r\rightarrow r e^{i\theta_r}$, $\alpha =2(\theta_q-\theta_p)$, $\beta =(\theta_r-\theta_p)$ and the rest of the parameters are already defined in Eqn.(\ref{prmtr}) and Eqn.(\ref{phi}).
\section{Breaking of the scaling ansatz}
Since the neutrino mass matrix obtained in Eqn.(\ref{8}), (\ref{9}) and (\ref{10}) are all invariant under scaling ansatz and thereby give rise to 
$\theta_{13}=0$ as well as $m_3=0$. Although vanishing value of $m_3$ is yet not ruled out however, the former, $\theta_{13}=0$ is refuted by the reactor experimental results. Popular paradigm is to consider $\theta_{13}=0$ at the leading order and by further perturbation nonzero value of $\theta_{13}$ is generated. We follow the same way to produce nonzero $\theta_{13}$ through small breaking of scaling ansatz. It is to be noted in our scheme, generation of nonzero $\theta_{13}$ necessarily needs breaking in $m_D$. To generate nonzero $m_3$ breaking in $\mu$ matrix is also necessary along with $m_D$, however, in Category B since $det(m_D=0)$ even after breaking in the $\mu$ matrix $m_\nu$ still gives one of the eigenvalue equal to zero. On the other hand for Category C and Category D, $\mu_2^4$ has always zero determinant because of being scaling ansatz invariant and therefore, leads to one zero eigenvalue as that of Category B.
It is the Category A for which we get nonzero $\theta_{13}$ as well as nonzero $m_3$ after breaking the scaling ansatz in both the matrices ($m_D$ and $\mu$).
\vskip 0.1in
\noindent
In the following, we invoke breaking of scaling ansatz in all four categories through\\
i) breaking in the Dirac sector ($\theta_{13}\neq0$, $m_3=0$)\\
ii) breaking in the Dirac sector as well as Majorana sector ($\theta_{13}\neq0$, $m_3\neq0$) and later we discuss separately both the cases.

\subsection{Breaking in the Dirac sector}
\subsubsection{Category A,B}

We consider minimal breaking of the scaling ansatz through a dimensionless real parameter $\epsilon$ in a single term of different $m_D$ matrices of those categories as
\bea
m_{D2}^4=\begin{pmatrix}
0&d_2&0\\
k_1(1+\epsilon) c_1&0&k_1 c_3\\
c_1&0&c_3
\end{pmatrix}
,m_{D6}^4=\begin{pmatrix}
0&0&d_3\\
k_1(1+\epsilon) c_1&k_1 c_2&0\\
c_1&c_2&0\\
\end{pmatrix}\label{11}
\eea
for Category A and
\bea
m_{D1}^4=\begin{pmatrix}
d_1&0&0\\
k_1 c_1&0&k_1(1+\epsilon) c_3\\
c_1&0&c_3\\
\end{pmatrix} ,
m_{D3}^4=\begin{pmatrix}
0&0&d_3\\
k_1(1+\epsilon) c_1&0&k_1 c_3\\
c_1&0&c_3\\
\end{pmatrix} \nonumber \\
m_{D4}^4=\begin{pmatrix}
d_1&0&0\\
k _1c_1&k_1(1+\epsilon) c_2&0\\
c_1&c_2&0\\
\end{pmatrix} ,
m_{D5}^4=\begin{pmatrix}
0&d_2&0\\
k_1(1+\epsilon) c_1&k_1 c_2&0\\
c_1&c_2&0\\
\end{pmatrix}
\eea
for Category B. We further want to mention that breaking considered in any 
element of the second row are all equivalent. For example, if we consider breaking in the `$23$' element of $m_{D2}^4$ it is equivalent to as considered in 
Eqn.(\ref{11}).  Neglecting the $\epsilon^2$ and higher order terms, the effective $m_\nu$ matrix comes out as
\bea
 m_\nu^{\prime AB\epsilon}&=& m_0\begin{pmatrix}
1& k_1p & p \\
k_1p  & k_1^2(q^2e^{i\theta}+p^2)& k_1(q^2e^{i\theta}+p^2)\\p & k_1(q^2e^{i\theta}+p^2)&(q^2e^{i\theta}+p^2)
\end{pmatrix}+m_0 \epsilon \begin{pmatrix}
0&0&0\\0&2k_1^2q^2e^{i\theta}&k_1q^2e^{i\theta}\\0&k_1q^2e^{i\theta}&0\\
\end{pmatrix}.\label{mAB}  
\eea
As mentioned earlier, that for Category B, $det(m_D)=0$ and it is not possible to generate $m_3\neq0$ even if we consider breaking in the $\mu$ matrices. On the other hand , the matrices in Category A posses $det(m_D)\neq0$ and thereby give rise to $m_3\neq0.$

 Now to calculate the eigenvalues, mixing angles, $J_{CP}$, the Dirac and Majorana phases we utilize the results obtained in ref.\cite{Adhikary:2013bma}, for a general complex matrix. We should mention that the formula obtained in ref.\cite{Adhikary:2013bma}, for Majorana phases is valid when all three eigenvalues are nonzero. However, when one of the eigenvalue is zero (in this case $m_3=0$) one has to utilize the methodology given in ref.\cite{sc2}, which shows, a general Majorana type mass matrix $m_\nu$  can be diagonalized as 
\bea
U^\dagger m_\nu U^*&=& diag(m_1,m_2,m_3)\eea
or alternely, \bea m_\nu &=& U diag(m_1,m_2,m_3)U^T \label{m1}\eea \\ where \bea U&=&U_{CKM} P_M.\eea\\
The mixing matrix $U_{CKM}$ is given by (following PDG\cite{Beringer:1900zz})convention)
\bea
U_{CKM}=
\begin{pmatrix}
c_{1 2}c_{1 3} & s_{1 2}c_{1 3} & s_{1 3}e^{-i\delta_{c p}}\\
-s_{1 2}c_{2 3}-c_{1 2}s_{2 3}s_{1 3} e^{i\delta_{c p} }& c_{1 2}c_{2 3}-s_{1 2}s_{1 3} s_{2 3} e^{i\delta_{c  p}} & c_{1 3}s_{2 3}\\
s_{1 2}s_{2 3}-c_{1 2}s_{1 3}c_{2 3}e^{i\delta_{c p}} & -c_{1 2}s_{2 3}-s_{1 2}s_{1 3}c_{2 3}e^{i\delta_{c p}} & c_{1 3}c_{2 3}
\end{pmatrix}
\eea
with  $c_{ij}=\cos \theta_{ij}$, $s_{ij} = \sin \theta_{ij}$ and $\delta_{CP}$ is the Dirac CP phase. The diagonal phase matrix $P_M$ is parametrized as
 \bea P_M& =& diag(1,e^{\alpha_M},e^{i(\beta_M +\delta_{CP})})\eea
 with $\alpha_M$ and $\beta_M+\delta_{CP}$ are the Majorana phases.

Writing Eqn.(\ref{m1}) explicitly with $m_3=0$ we can have expressions for six independent elements of $m_\nu$ in terms of the mixing angles, two eigenvalues and the Dirac CP phase, from which the $m_{11}$ element can be expressed as
\bea
 m_{11} & =& c_{12}^2 c_{13}^2m_1 +s_{12}^2 c_{13}^2m_2 e^{2i\alpha_M}\label{m2}
 \eea
 and therefore the Majorana phase $\alpha_M$ comes out as
 \bea
 \alpha_M &=& \frac{1}{2}\cos^{-1}\left\lbrace \frac{|m_{11}|^2•}{2c_{12}^2s_{12}^2c_{13}^4m_1m_2•}-\frac{(c_{12}^4m_1^2+s_{12}^4m_2^2)}{•2c_{12}^2s_{12}^2m_1m_2}\right\rbrace. \label{m3}\eea
 The Jarlskog measure of CP violation $J_{CP}$ is defined in usual way as
 \bea 
 J_{CP}&=& \frac{Im(h_{12}h_{23}h_{31})}{(\Delta m_{21}^2)(\Delta m_{32}^2)(\Delta m_{31}^2)} 
 \eea
 where $h$ is a hermitian matrix constructed out of $m_\nu$ as $h=m_\nu m_\nu^{\dagger}$.
 \subsubsection{Category C}

In this case breaking is considered in $m_D$ as 
\bea
m_{D1}^4=\begin{pmatrix}
d_1&0&0\\
k_1(1+\epsilon) c_1&k_1 c_2&0\\
c_1&c_2&0\\
\end{pmatrix},
m_{D4}^4 =\begin{pmatrix}
d_1&0&0\\
k_1(1+\epsilon) c_1&0&k_1 c_3\\
c_1&0&c_3\\
\end{pmatrix}
\eea
and the scaling ansatz broken $m_\nu$ appears as  \bea
m_\nu^{\prime C \epsilon} = m_0 \begin{pmatrix}
 1&k_1(p+qe^{i\theta})&p+qe^{i\theta}\\
 k_1(p+qe^{i\theta})&k_1^2(2pqe^{i\theta}+p^2)&k_1(2pqe^{i\theta}+p^2)\\
 p+qe^{i\theta}&k_1(2pqe^{i\theta}+p^2)&(2pqe^{i\theta}+p^2)\\
 \end{pmatrix} \nonumber \\+m_0 \epsilon \begin{pmatrix}
 0&k_1qe^{i\theta}&0\\k_1qe^{i\theta}&2k_1^2pqe^{i\theta}&k_1pqe^{i\theta}\\
 0&k_1pqe^{i\theta}&0\\
 \end{pmatrix}.
 \eea
\subsubsection{Category D} 

Breaking in $m_D$ in this case is incorporated through
\bea
m_{D2}^4=\begin{pmatrix}
0&d_2&0\\
k_1 c_1&0&k_1(1+\epsilon) c_3\\
c_1&0&c_3\\
\end{pmatrix},
m_{D3}^4=\begin{pmatrix}
0&0&d_3\\
k_1 c_1&0&k_1(1+\epsilon) c_3\\
c_1&0&c_3\\
\end{pmatrix} \nonumber \\
m_{D5}^4=\begin{pmatrix}
0&d_2&0\\
k_1 c_1&k_1(1+\epsilon) c_2&0\\
c_1&c_2&0\\
\end{pmatrix},
m_{D6}^4=\begin{pmatrix}
0&0&d_3\\
k_1 c_1&k_1(1+\epsilon) c_2&0\\
c_1&c_2&0\\
\end{pmatrix}
\eea
and the corresponding $m_\nu$ comes out as
\bea
m_\nu^{\prime D \epsilon} =m_0\begin{pmatrix}
0&k_1p&p\\
k_1p&k_1^2(q^2e^{i\alpha}+2rpe^{i\beta})&k_1(q^2e^{i\alpha}+2rpe^{i\beta})\\
p&k_1(q^2e^{i\alpha}+2rpe^{i\beta})&(q^2e^{i\alpha}+2rpe^{i\beta})
\end{pmatrix} \nonumber \\ +m_0\epsilon  \begin{pmatrix}
0&0&0\\0&2k_1^2rpe^{i\beta}&k_1rpe^{i\beta}\\0&k_1rpe^{i\beta}&0\\
\end{pmatrix}.
\eea
 \subsection{Numerical Analysis}
 In order to perform the numerical analysis to obtain allowed parameter space
 we utilize the neutrino oscillation data obtained from global fit shown in \textbf{Table \ref{tl}}.
 \begin{table}[!h]
 \caption{Input experimental values\cite{Tortola:2012te}}\label{tl}
 \begin{center}
 \begin{tabular}{|c|c|}
\hline 
Quantity & 3$\sigma$ ranges \\ 
\hline 
$|\Delta m_{31}^2|$ N& 2.31$< \Delta m_{31}^2(10^3 eV^{-2})<2.74$ \\ 
\hline 
$|\Delta m_{31}^2|$ I& 2.21$< \Delta m_{31}^2(10^3 eV^{-2})<2.64$ \\ 
\hline 
 $\Delta m_{21}^2$& 7.21$< \Delta m_{21}^2(10^5 eV^{-2})<8.20$ \\ 
\hline 
$\theta_{12}$ & $31.3^o<\theta_{12}<37.46^o$ \\ 
\hline 
$\theta_{23}$ &  $36.86^o < \theta_{23}<55.55^o$  \\  
\hline 
$\theta_{13}$ &  $7.49^o < \theta_{13}< 10.46^o$  \\ 
\hline 
\end{tabular} 
\end{center} 
 \end{table}
 \subsubsection{Category A,B}
 We first consider Category A,B for which the neutrino mass matrix is given in Eqn.(\ref{mAB}). The parameter $\epsilon$ is varied freely to fit the extant data and it is constrained as $0.04<\epsilon<0.7$. However, to keep the ansatz breaking effect small we restrict the value of  $\epsilon$ only upto 0.1. For this range of $\epsilon$ ($0<\epsilon<0.1$) under consideration the parameter spaces are obtained as $1.78<p<3.40$, $1.76<q<3.42$ and $0.66<k_1<1.3$.  It is interesting to note a typical feature of this category is that the Dirac CP phase $\delta_{CP}$ comes out too tiny  and thereby generating almost vanishing value of $J_{CP}$ ($\approx 10^{-6}$) while the range of the only Majorana phase in this category is obtained as $77^o<\alpha_M<90^o$. \\
 \begin{figure}[h!]
 \includegraphics[scale=.6]{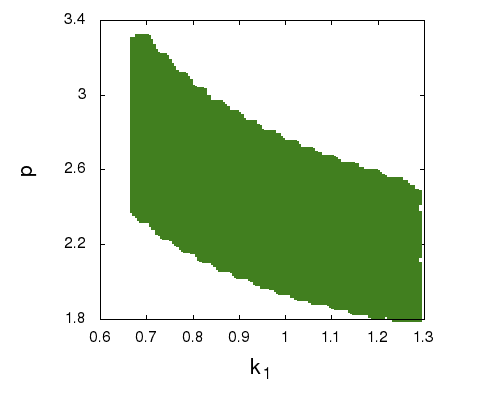}          \includegraphics[scale=.6]{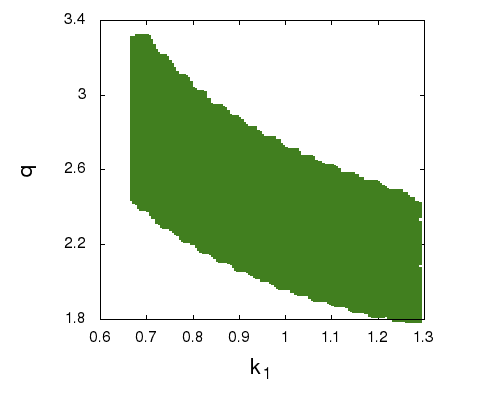}\\
\caption{Plot of $p$ vs $ k_1$ (left), $q$ vs $k_1$ (right) for the Category A,B with $\epsilon=0.1.$}\label{fig1}
\end{figure}

\noindent

\begin{figure}[!h]
\begin{center}
\includegraphics[scale=.6]{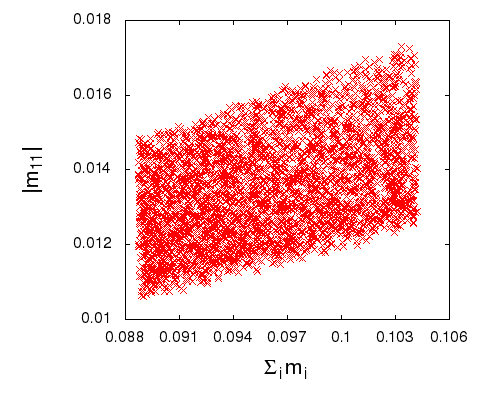}\\
\caption{ Plot of $|m_{11}|$ vs $\Sigma_i m_i$ for Category  A,B with $\epsilon=0.1$.}\label{fig2}
\end{center}
\end{figure}
\noindent
As one of the eigenvalue $m_3=0$ therefore, the hierarchy of the masses is clearly inverted in this category. The sum of the three neutrino masses $\Sigma_i m_i(=m_1+m_2+m_3)$ and $|m_{11}|$ are obtained as  0.088 eV $<\Sigma_i m_i<$ 0.104 eV and 0.0102 eV $<|m_{11}|<$ 0.0181 eV which predict the value of the two quantities below the present experimental upper bounds. To illustrate the nature of variation, in figure \ref{fig1} we plot $p$ vs $k_1$ and $q$ vs $k_1$ while in figure \ref{fig2} a correlation plot of $\Sigma_i m_i$ with $|m_{11}|$ is shown for $\epsilon=0.1$ and it is also seen from figure \ref{fig1} and  \ref{fig2} that the ranges of the parameters do not differ much compare to the values obtained for the whole range of $\epsilon $ parameter. \\

\noindent
In brief, distinguishable characteristics of this category are i) tiny $J_{CP}$ and $\delta_{CP}$ ii) inverted hierarchy of the neutrino masses. At the end of this section we will further discuss the experimental testability of these quantities for all the categories.
\subsubsection{Category C}
In this case it is found that a small breaking of $\epsilon$ ($0.02<\epsilon<0.09$) is sufficient to accommodate all the oscillation data. We explore the parameter space and the ranges obtained as $3.42<p<6.07$, $1.68<q<3.02$ and $0.7<k_1<1.32$. The hierarchy obtained in this case is also inverted due to the vanishing value of $m_3$. 
\begin{figure}[!h]
\includegraphics[scale=.6]{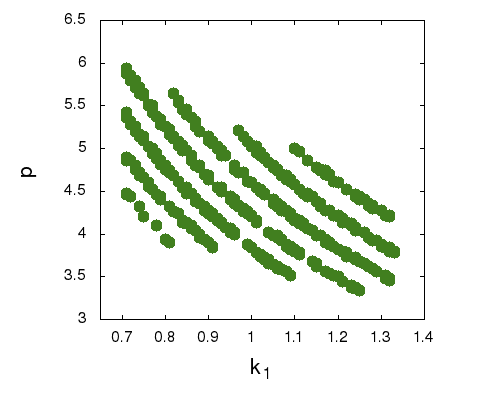}   \includegraphics[scale=.6]{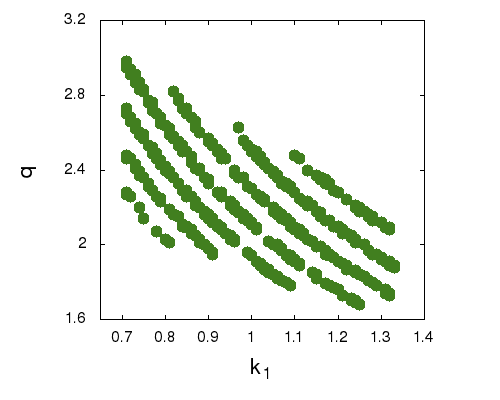} \\
\caption{ Plot of $p$ vs $ k_1$ (left), $q$ vs $k_1$ (right) for the Category C with $\epsilon=0.09.$}\label{fig3}
\end{figure}
 The other two quantities $\Sigma_i m_i$ and  $|m_{11}|$ come out as 0.0118 eV $<|m_{11}|<$ 0.019 eV and 0.088 eV $<\Sigma_i m_i<$ 0.105 eV.  Similar to the previous category  $J_{CP}$ is vanishingly small due to low value of $\delta_{CP}$.  The range of the  Majorana phase $\alpha_M$ is obtained as $81^o<\alpha_M<89^o$. In figure \ref{fig3} we plot $k_1$ vs $p$ and $k_1$ vs $q$ for  $\epsilon=0.09$ that predicts almost the same ranges of the parameters ($p$, $q$ and $k_1$) and all other quantities ($|m_{11}|$,  $\Sigma_i m_i$, $\alpha_M$ and $J_{CP}$) as  obtained from the whole range of $\epsilon$. We present a correlation plot of $\Sigma_i m_i$ with  $|m_{11}|$ in figure \ref{fig4}.
\begin{figure}[h!]
 \begin{center}
  \includegraphics[scale=.6]{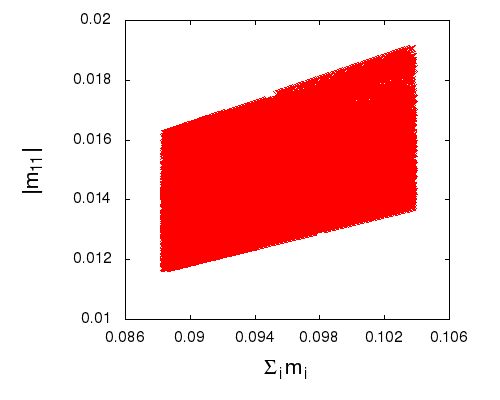}\\
  \caption{Plot of $|m_{11}|$ vs $\Sigma_i m_i$ for Category  C with $\epsilon=0.09.$}\label{fig4}
  \end{center}
  \end{figure}
\subsubsection{Category D}\label{s1}
In case of Category D, although a priori it is not possible to rule out $m_\nu^{\prime D\epsilon}$ without going into the detailed numerical analysis, however in this case even if with $\epsilon=1$ it is not possible to accommodate the neutrino oscillation data. Specifically, the value of $\theta_{13}$ is always beyond the reach of the parameter space. Exactly for the same reason the $m_\nu$ matrix of type $A_3$ in Eqn.(\ref{rldout}) is phenomenologically ruled out.
\subsection{Breaking in Dirac+Majorana sector}
In this section we  focus on the phenomenology of the neutrino mass matrix where the scaling ansatz is broken in both  the sectors. This type of breaking is only relevant for Category A since in this case $m_D$ is nonsingular after  breaking of the ansatz and the resultant $m_\nu$ gives rise to nonzero $\theta_{13}$ along with $m_3\neq0$. In all the other categories due to the singular nature of $m_D$, inclusion of symmetry breaking in the Majorana sector will not generate $m_3\neq0$. Thus we consider only Category A under this scheme.
\subparagraph{}
We consider the breaking in $m_D$ as mentioned in Eqn.(\ref{11}) and the ansatz broken texture of $\mu_1^4$ matrix is given by
\bea
\mu_1^4 &=& \begin{pmatrix}
r_1&0&0\\
0&k_2^2s_3&k_2 (1+\epsilon^\prime)s_3\\
0&k_2(1+\epsilon^\prime) s_3&s_3\\
\end{pmatrix} 
\eea
where $\epsilon^\prime$ is a dimensionless real parameter. The effective neutrino mass matrix $m_\nu$ comes out as 
\bea
 m_{\nu \epsilon ^ \prime}^{\prime A \epsilon}= m_0\begin{pmatrix}
1& k_1p & p \\
k_1p  & k_1^2(q^2e^{i\theta}+p^2)& k_1(q^2e^{i\theta}+p^2)\\p & k_1(q^2e^{i\theta}+p^2)&(q^2e^{i\theta}+p^2)
\end{pmatrix}+m_0 \epsilon \begin{pmatrix}
0&0&0\\0&2k_1^2q^2e^{i\theta}&k_1q^2e^{i\theta}\\0&k_1q^2e^{i\theta}&0\\
\end{pmatrix} \nonumber \\ +m_0 \epsilon^\prime \begin{pmatrix}
0&k_1p&p\\k_1p&0&0\\p&0&0
\end{pmatrix}.
\eea
\subsubsection{Numerical results}
\begin{figure}[h!]
\begin{center}
\includegraphics[scale=.6]{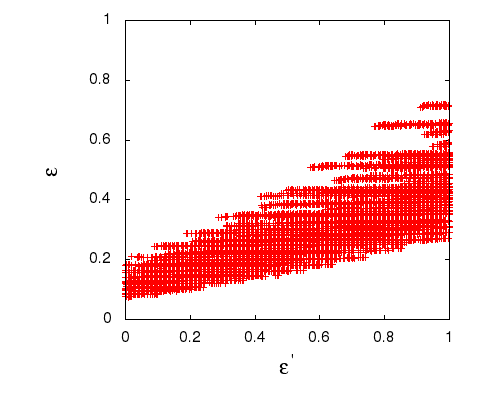}
\caption{Correlated plot of $\epsilon$ with $\epsilon^{\prime}$.}\label{fig5}
\end{center}
 \end{figure}

\begin{figure}[h!]
\includegraphics[scale=.6]{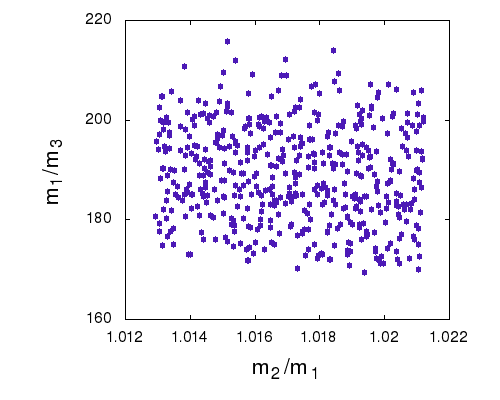}        \includegraphics[scale=.6]{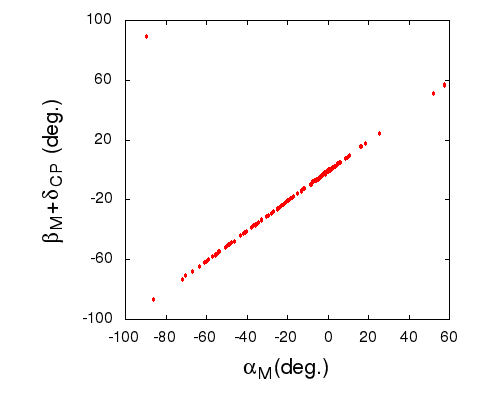}\\
 \caption{ Plot of $(m_1/m_3)$ vs $ (m_2/m_1)$ (left) and $\beta_M+\delta_{CP}$ vs $\alpha_M$ (right) after breaking of the scaling ansatz in both the sectors of Category A for a representative value of $\epsilon=\epsilon^{\prime}=0.1$.}\label{fig6}
\end{figure}

As mentioned above, $\epsilon^{\prime}=0$ leads to inverted hierarchy with $m_3=0$ and thus to generate nonzero $m_3$ a small value of  $\epsilon^{\prime}$ is needed. Similar to the previous cases two breaking parameters $\epsilon$ and $\epsilon^{\prime}$ can be varied freely through the ranges that are sensitive to the oscillation data and are obtained as $0.06<\epsilon<0.68$ and $0<\epsilon^{\prime}<1$. It is to be noted that although the $\epsilon$ parameter is restricted due to $\theta_{13}$ value, $\epsilon^{\prime}$ is almost insensitive to $\theta_{13}$ and it can vary within a wide range as $0<\epsilon^{\prime}<1$.  A correlation plot of $\epsilon$ with $\epsilon^{\prime}$ is shown in figure \ref{fig5}. However, as mentioned  earlier, the effect of the breaking term should be smaller than the unbroken one, therefore, to obtain the parameter space for this category we consider breaking of the scaling ansatz in both the sectors only upto 10 \% and consequently for all combinatorial values of $\epsilon$ and $\epsilon^{\prime}$ the parameters $p$, $q$ and $k_1$  vary within the ranges as  $1.07<p<3.10$, $1.03<q<3.12$ and $0.67<k_1<1.31$. Interestingly, although all the eigenvalues are nonzero in this case, the hierarchy is still inverted. $J_{CP}$ is found to be tiny ($\approx 10^{-6}$) again due to small value of $\delta_{CP}$. The Majorana phases are obtained as $-96^o<\alpha_M<74^o$ and $-100^0<\beta_M+\delta_{CP}<102^o$ followed by  the bounds on $\Sigma_im_i$ and $|m_{11}|$ as 0.088 eV $<\Sigma_i m_i<$ 0.11 eV and 0.010 eV $<|m_{11}|<$ 0.022 eV which are well below the present experimental upper bounds. In figure \ref{fig6} we demonstrate the above predictions for $\epsilon=\epsilon^{\prime}=0.1$. In the left panel of figure \ref{fig6} the inverted hierarchical nature is shown and in the right panel variation of the Majorana phases is demonstrated.  
  \\
\\
\noindent
Some comments are in order regarding predictions of the present scheme:\\
\noindent  
  1. After precise determination of $\theta_{13}$ taking full account of reactor neutrino experimental data, it is shown that the hierarchy of the light neutrino masses can be probed through combined utilization of NO$\nu$A and T2K\cite{Ieki:2014bca} neutrino oscillation experimental results in near future. Thus the speculation of hierarchy in the present scheme will be clearly 
verified. Moreover, taking the difference of probabilities between $P(\nu_\mu \rightarrow \nu_e)$ and $P(\bar{\nu_\mu}\rightarrow \bar{\nu_e})$ information on the value of $J_{CP}$ can be obtained using neutrino and anti neutrino beams.\\
\noindent  
2. More precise estimation of the sum of the three light neutrino masses will be obtained utilizing a combined analysis with PLANCK data\cite{Ade:2013zuv} and other cosmological and astrophysical experiments\cite{Lesgourgues:2014zoa} such as, Baryon oscillation spectroscopic survey, The Dark energy survey, Large Synoptic Survey Telescope or the Euclid satellite data\cite{1475-7516-2013-01-026} etc. Such type of analysis will push $\Sigma_i m_i\sim 0.1$ eV (at the $4\sigma$ level for inverted ordering) and $\Sigma_i m_i\sim 0.05$ eV (at the $2\sigma$ level for normal ordering). Thus the prediction of the value of $\Sigma_i m_i$ in the different categories discussed in the present work 
will also be tested in the near future. Furthermore, the NEXT-100\cite{DavidLorcafortheNEXT:2014fga} will probe the value of $|m_{11}|$ up to $0.1$ eV which is a more precise value than the EXO-200\cite{Auger:2012ar} experimental range (0.14-0.38 eV).

\section{Summary and conclusion}

In this work we explore the phenomenology of neutrino mass matrix obtained due to inverse seesaw mechanism adhering
i) Scaling ansatz, ii) Texture zeros within the framework of $SU(2)_L \times U(1)_Y$ model with three right handed neutrinos and three left chiral singlet 
fermions.
Throughout our analysis we choose a basis in which the charged lepton mass matrix ($m_E$) and the $M_{RS}$ matrix (appeared in inverse seesaw mechanism due to the coupling of $\nu_R$ and $S_L$) are  diagonal. It is found that four is the maximum number of zeros that can be allowed in $m_D$ and $\mu$ matrices to obtain viable phenomenology. We classify different four zero textures in four different categories depending upon their generic form. Since scaling ansatz invariance always gives rise to $\theta_{13}=0$, we have to break such ansatz. We consider breaking in $m_D$ and also in $\mu$ matrices. We explore the parameter space and it is seen that one category (Category D) is ruled out phenomenologically. The hierarchy obtained in all the cases is inverted and it is interesting to note that all such categories give rise to tiny $CP$ violation measure $J_{CP}$ due to small value of $\delta_{CP}$. In conclusion, further observation of hierarchy of neutrino masses and CP violation in the leptonic sector in the forthcoming experiments will conclusively refute or admit  all these categories obtained in the present scheme.\\\\\\

\noindent
\textbf{ Acknowledgement}\\
We thank Mainak Chakraborty for discussion and computational help.

\end{document}